\pdfoutput=1
\documentclass[%
superscriptaddress,
twocolumn,
aps,
prb,
]{revtex4-2}

\usepackage{graphicx}
\usepackage{dcolumn}
\usepackage{bm}
\usepackage{amsmath, esint}
\usepackage{amssymb}
\usepackage{physics}
\usepackage{bigints}
\usepackage{comment}
\usepackage{xcolor}
\renewcommand{\vec}{\vectorbold}

\DeclareMathAlphabet\mathbfcal{OMS}{cmsy}{b}{n}

\newcommand{\PRB}[1]{{\textcolor{black}{#1}}}

\newcommand{\ofx}{(\vec{x})}

\usepackage{hyperref}

\hypersetup{
    colorlinks,
    citecolor=black,
    filecolor=black,
    linkcolor=black,
    urlcolor=black
}

\begin{document}

\preprint{APS/123-QED}

\title{Negative electrohydrostatic pressure between superconducting bodies}

\author{Thomas J. Maldonado}
\email{maldonado@princeton.edu}
\affiliation{%
Department of Electrical and Computer Engineering, Princeton University, Princeton, NJ 08544, USA}
\author{Dung N. Pham}%
\affiliation{%
Department of Electrical and Computer Engineering, Princeton University, Princeton, NJ 08544, USA}%
\author{Alessio Amaolo}
\affiliation{Department of Chemistry, Princeton University, Princeton, NJ 08544, USA}
\author{Alejandro W. Rodriguez}
\affiliation{%
Department of Electrical and Computer Engineering, Princeton University, Princeton, NJ 08544, USA}%
\author{Hakan E. Türeci}
\affiliation{%
Department of Electrical and Computer Engineering, Princeton University, Princeton, NJ 08544, USA}%

\date{\today}
\begin{abstract}
    By applying a hydrodynamic representation of non-relativistic scalar electrodynamics to the superconducting order parameter, we predict a negative (attractive) pressure between planar superconducting bodies. For conventional superconductors with London penetration depth \(\lambda_\text{L} \approx 100 \text{ nm}\), the pressure reaches tens of \(\text{N/mm}^2\) at angstrom separations. The resulting surface energies are in better agreement with experimental values than those predicted by the Hartree-Fock theory, and \PRB{the emergent electric-field screening length is comparable to that of the Thomas-Fermi theory. The model} circumvents the bulk limitations of the Bardeen-Cooper-Schrieffer and Ginzburg-Landau theories to the analysis of superconducting quantum devices.
\end{abstract}

\maketitle

\PRB{\section{Introduction}}
In conventional superconductors, steady-state bulk phenomena are accurately described by both the Bardeen-Cooper-Schrieffer (BCS)~\cite{BCS} and Ginzburg-Landau (GL)~\cite{GL} theories. The former provides a microscopic origin for superconductivity via the phonon-mediated pairing of electrons into bosonic quasiparticles known as Cooper pairs, while the latter provides a phenomenological description of the resulting condensate~\cite{bec} with a macroscopic order parameter representing its mean-field wave function. The two theories were shown to be equivalent near the superconducting critical temperature~\cite{gorkov}, and both reproduce the London theory~\cite{london}. Though the BCS theory is sufficiently general to predict time-dependent bulk phenomena, an effective macroscopic theory is desirable when such effects are triggered by electromagnetic sources in spatially inhomogeneous domains. To this end, generalized GL equations have been proposed to capture boundary and wave effects present in complex geometries~\cite{oripov2020time}, but a consensus has not been reached on their validity far below the critical temperature, a regime all too familiar to the burgeoning area of superconducting quantum devices~\cite{clarke2008superconducting}. 

In this letter, we present and explore predictions offered by a hydrodynamic representation of non-relativistic scalar electrodynamics applied to the superconducting order parameter at zero temperature. Few attempts have been made to solve this model's equations of motion (EOM) exactly~\cite{dec-qed}, but simplified versions have been considered via relaxations of minimal coupling~\cite{feynman, greiter1989hydrodynamic, PhysRevLett.64.587, salasnich2009hydrodynamics} and can be credited as the underpinning of Josephson phenomena and circuit quantum electrodynamics~\cite{RevModPhys.93.025005}. Such approximate descriptions of light-matter interactions have enabled coveted numerical analyses of superconducting circuits embedded in electromagnetic resonant structures~\cite{PhysRevLett.108.240502, minev2021energy} \PRB{far below the critical temperature}, but they rely on London-like boundary conditions between superconducting and non-superconducting domains that seem to harbor serious inconsistencies~\cite{PhysRevB.69.214515}. Our goal is not to provide a rigorous derivation of the theory (the literature contains some attempts~\cite{aitchison1995effective, ao1995nonlinear}), but rather to demonstrate that its unapproximated form circumvents \PRB{the aforementioned} spatial partitioning and implies a pressure between planar superconducting bodies that can be measured to determine its validity.

\begin{figure*}
    \centering
    \includegraphics[width=\textwidth,height=\textheight,keepaspectratio]{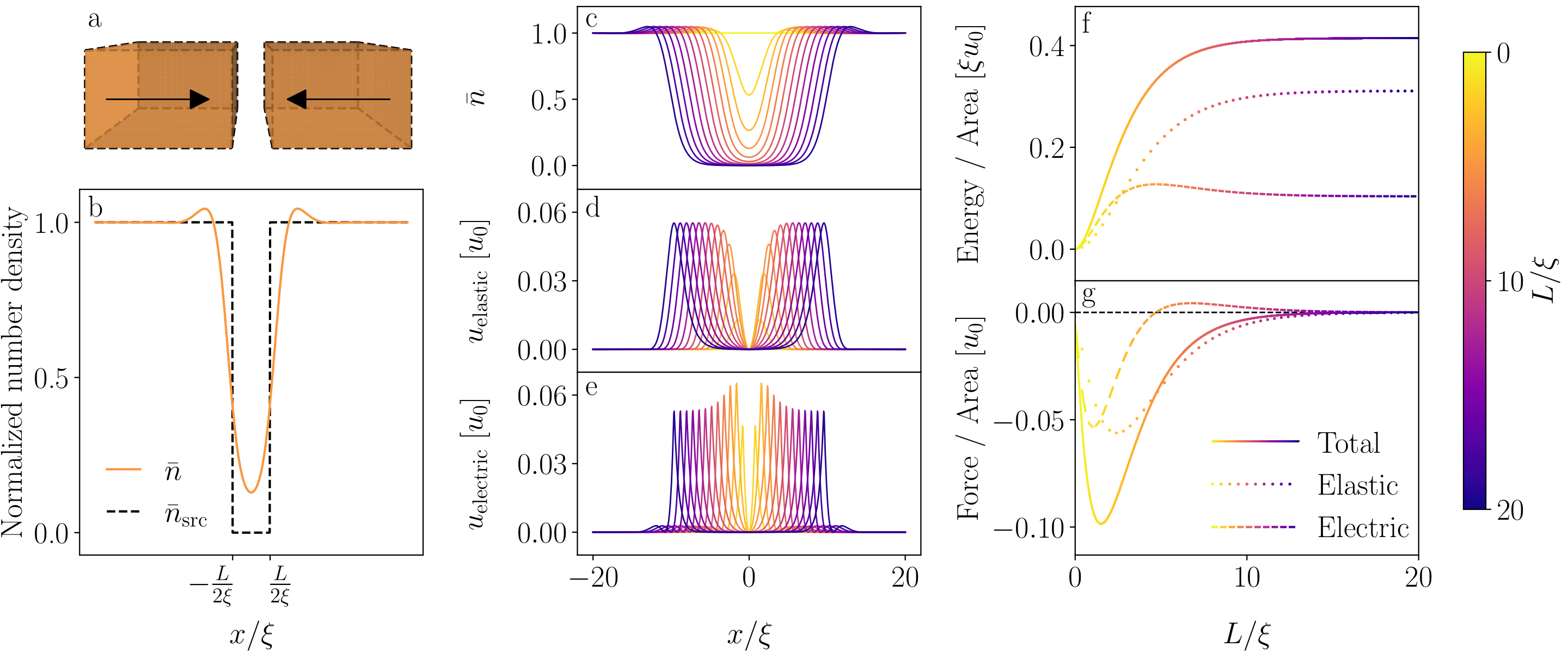}
    \caption{The main result is summarized by a free body diagram (a) depicting the attractive force between two planar superconducting bodies. Calculation of this pressure begins with a numerical solution \(\bar{n}\) to the electrohydrostatic condition sourced by two finitely separated ionic backgrounds \(\bar{n}_\text{src}\). An example solution with separation length \(L = 4.8\xi\) is depicted in (b). We next calculate the electrostatic distribution of the normalized number density (c), elastic energy density (d), and electric energy density (e) for a range of separation lengths specified by the color bar. Energy densities are plotted in units of \(u_0 = \hbar mc/\left(\mu_0q^2\lambda_\text{L}^3\right)\) and are spatially integrated to find the energy per unit area (f) as a function of the separation length, whose negative derivative with respect to \(L\) yields the pressure (g). We note that the electric pressure changes sign at \(L \approx 4.8\xi\).}
    \label{fig:main_fig}
\end{figure*}

While our model shares similarities with the GL theory in that it describes the superconducting condensate with an order parameter, it differs in at least four important ways. First, in contrast to the diffusive time-dependent GL equations, our model entails wave-like dynamics implied by Schrödinger's equation. Second, we employ minimal coupling to all electromagnetic degrees of freedom, including the electric field via Gauss's law and Maxwell's correction to Ampere's law. Third, we incorporate arbitrary arrangements of both external drives and ionic backgrounds via normal (non-superconducting) source distributions. We take the latter to be static in nature, akin to the Jellium model of a metallic conductor~\cite{RevModPhys.65.677}, but generalizable to include dynamical fluctuations for effective descriptions of phononic excitations. Fourth, in considering regimes far below the critical temperature, we omit the self-interaction term that governs the GL phase transition. In our model, nonlinear phenomena arise instead from our more general treatment of light-matter interactions, and the Higgs mechanism that yields the condensate's equilibrium number density via spontaneous symmetry breaking of the \(U(1)\) gauge group is replaced by requirement from the EOM that the \PRB{entire system exhibits charge neutrality. Upon explicit symmetry breaking by the free currents, this final property yields a bulk superconducting charge density equal and opposite to the ionic background and has been studied in certain astrophysical settings \cite{charged_condensation}}.

\PRB{The paper is organized in the following manner. We formally define the model in Section \ref{model_statement} via presentation of the Lagrangian, EOM, and Hamiltonian. In Section \ref{results}, we enumerate the results of our electrostatic analysis, beginning with an equation of state satisfied by solutions in electrohydrostatic equilibrium. We then solve this electrohydrostatic condition numerically in the context of two planar superconducting bodies separated by vacuum. By considering variations in the system's electrohydrostatic energy with respect to the separation length, we find a negative (attractive) pressure between the two bodies that peaks at an emergent healing length. We proceed by calculating the resulting surface energies for a number of elemental superconductors and find better agreement with experimental values than the Hartree-Fock (HF) method applied to the normal fermionic state. Finally, we compare our model's screening response to that of a normal metal. While our derived response is functionally distinct from that of the Thomas-Fermi (TF) theory, both yield remarkably similar values for the electric-field screening length in elemental superconductors. We conclude the paper in Section \ref{outlook} with a discussion of the other forces relevant to a measurement of the electrohydrostatic pressure, along with an outlook toward future dynamical analyses.}

\PRB{\section{Model statement}\label{model_statement}} 
Throughout the text, we employ the covariant formulation of electromagnetism with the Minkowski metric \(\eta^{\mu\nu} = \text{diag}(+,-,-,-)^{\mu\nu}\), and we refer to the components of a four-vector as \mbox{\(X^\mu \equiv \left(X_0, \vec{X}\right)^\mu\)}. Though the model describes non-relativistic charged superfluids, we find that a relativistic notation provides useful physical insight. We assume the effective Lagrangian governing the evolution of the order parameter \(\psi \equiv \sqrt{n}e^{i\theta}\) and the electromagnetic four-potential \(A^\mu\) is given by the non-relativistic theory of scalar electrodynamics under minimal light-matter coupling,

\begin{equation}\label{eq:lagrangian}
    \begin{aligned}
        \mathcal{L} &= \psi^*\left(i\hbar\frac{\partial}{\partial t} - qcA_0 - \frac{1}{2m}\left(\frac{\hbar}{i}\grad - q\vec{A}\right)^2\right)\psi\\
        & \qquad \qquad \qquad -\frac{1}{4\mu_0}F^{\mu\nu}F_{\mu\nu} - A^\mu j_\mu,
    \end{aligned}
\end{equation}
where \(F^{\mu\nu} \equiv \partial^\mu A^\nu - \partial^\nu A^\mu\) is the electromagnetic tensor, \(j^\nu\) is the four-current generated by normal charges, and \(q\) and \(m\) are the charge and mass of the superconducting charge carriers, respectively. The EOM arising from this Lagrangian couple Maxwell's equations for the four-potential and Schrödinger's equation for the order parameter,
\begin{subequations} \label{eq:maxwell_schrodinger}
    \begin{gather}
        \partial_\mu F^{\mu\nu} = \mu_0 \left(\mathcal{J}^\nu + j^\nu\right) \\
        i\hbar\dot{\psi} = \left(\frac{1}{2m}\left(\frac{\hbar}{i}\grad - q\vec{A}\right)^2 + qcA_0\right)\psi,
    \end{gather}
\end{subequations}
where \mbox{\(\mathcal{J}^\nu \equiv qn\left(c, \vec{v}\right)^\nu\)} is the four-current generated by superconducting charges with number density \(n\) and flow velocity \mbox{\(\vec{v} \equiv \left(\hbar\grad{\theta} - q\vec{A} \right)/m\)}. As derived in \PRB{Appendix \ref{electrohydrodynamic_hamiltonian_appendix}}, the system's Hamiltonian can be expressed in an electrohydrodynamic form as 
\begin{equation}\label{eq:hamiltonian}
    \mathcal{H} = \frac{\epsilon_0}{2} \abs{\vec{E}}^2 + \frac{1}{2\mu_0}\abs{\vec{B}}^2 + n\left(\frac{1}{2}mv^2\right) + \frac{\hbar^2}{8m}n\abs{\grad \ln n}^2,
\end{equation}
with \(\vec{E} = -c\grad A_0 - \dot{\vec{A}}\) the electric field, \mbox{\(\vec{B}=\curl\vec{A}\)} the magnetic field, \(n\) the superconducting number density, and \(v \equiv \abs{\vec{v}}\) the flow speed. Eq.~\eqref{eq:hamiltonian} represents a decomposition of the total energy density into electric, magnetic, kinetic, and elastic components, respectively~\cite{fuchs2011lagrangian}. \PRB{Before proceeding, we emphasize that unlike Josephson junction models based on two order parameters, Eq.~\eqref{eq:lagrangian} assumes a single order parameter defined over all space.}

\begin{table*}[hbt!] \label{table}
     \begin{tabular}{ |c|c|c|c|c|c|c|c|c| } 
        \hline
        Element & \(r_\text{s}\) [\(a_0\)] & \(\lambda_\text{L}\) [nm] & \(\sigma_\text{n}\) [\(\text{J}/\text{m}^2\)] & \(\sigma_\text{sc}\) [\(\text{J}/\text{m}^2\)] & \(\sigma_\text{expt}\) [\(\text{J}/\text{m}^2\)] & \(\xi_\text{TF}\) [\AA] & \(\xi\) [\AA]\\
        \hline
        Al & 2.07 & 16 & -0.73 & 0.90 & 1.2 & 0.49 & 0.39\\ 
        In & 2.41 & 20.5 & -0.012 & 0.48 & 0.70 & 0.53 & 0.44\\ 
        Sn & 2.22 & 35 & -0.28 & 0.13 & 0.70 & 0.51 & 0.58\\ 
        Pb & 2.30 & 30.5 & -0.13 & 0.18 & 0.60 & 0.51 & 0.54\\ 
        \hline
    \end{tabular}
    \caption{Surface energies and screening lengths are calculated for aluminum, indium, tin, and lead at zero temperature. Fermionic models based on the Hartree-Fock method employ the metal's Wigner-Seitz radius \(r_\text{s}\)~\cite{ashcroft2022solid} in the normal state to predict the surface energy \(\sigma_\text{n}\) \PRB{(obtained via interpolation of the data in~\cite{lang1970theory})} and Thomas-Fermi screening length \(\xi_\text{TF}\) \cite{ashcroft2022solid}. Our model employs the metal's London penetration depth \(\lambda_\text{L}\) (Al \cite{BCS}, In \cite{dheer1961surface}, Sn \cite{BCS}, Pb \cite{PhysRevB.2.2519}) in the superconducting state to predict the surface energy \(\sigma_\text{sc}\) given by Eq.~\eqref{eq:surface_energy}, and healing length \(\xi\) given by Eq.~\eqref{eq:healing_length}. Experimental values for the surface energy \(\sigma_\text{expt}\) were averaged over the data provided by~\cite{vitos1998surface}.}
    \label{table:surface_energies}
\end{table*}

\PRB{\section{Results}\label{results}}
\PRB{\subsection{Equation of state}}
We now limit our focus to electrostatic systems, which are recovered by enforcing that all currents vanish \mbox{\(\mathbfcal{J} = \vec{j} = \vec{0}\)}. We first introduce the bulk superconducting number density \(n_\text{s}\) and two important length scales: the London penetration depth \(\lambda_\text{L} = \sqrt{m/\left(\mu_0 q^2 n_\text{s}\right)}\) and the Compton wavelength \(\lambda_\text{C} = h/(mc)\). In terms of the normalized number densities \mbox{\(\bar{n} \equiv n/n_\text{s} = \mathcal{J}_0/(cqn_\text{s})\)} and \mbox{\(\bar{n}_\text{src} \equiv - j_0/(cqn_\text{s})\)}, Eqs.~\eqref{eq:maxwell_schrodinger} reduce to the electrohydrostatic condition,
\begin{equation} \label{eq:electrostatics}
    \bar{n} + 2\xi^4\laplacian\frac{\laplacian \sqrt{\bar{n}}}{\sqrt{\bar{n}}} = \bar{n}_\text{src},
\end{equation}
revealing the established \cite{gabadadze2009effective} healing length \(\xi\) given by
\begin{equation} \label{eq:healing_length}
    \xi \equiv \sqrt{\frac{\lambda_\text{L}\lambda_\text{C}}{4\pi}}.
\end{equation}
As shown in \PRB{Appendix \ref{electrohydrostatic_condition_appendix}}, Eq.~\eqref{eq:electrostatics} is a self-consistent statement of Gauss's law that expresses the balance between electric and elastic forces in the electrostatic distribution of the fluid: \(q\vec{E} = \grad Q\) with \mbox{\(Q= - \hbar^2\left(\laplacian\sqrt{n}/\sqrt{n}\right)/(2m)\)} the well-known quantum potential~\cite{PhysRev.85.166}. Because of the nonlinear term, proving the existence or
uniqueness of solutions \(\bar{n}\) is nontrivial and remains an open
problem. We may nonetheless make some qualitative observations regarding solutions
to Eq.~\eqref{eq:electrostatics}. First, we anticipate the
asymptotic behavior \(\bar{n} \to \bar{n}_\text{src} = 1\) in the
bulk. Second, spatial derivatives in the nonlinear term ensure \(C^4\)
continuity of \(\bar{n}\) over all spatial coordinates. To avoid
introducing additional length scales, we focus here on
piecewise-constant sources \(\bar{n}_\text{src}\) that take values zero outside and one inside the superconducting material. 

\PRB{\subsection{Pressure}}
To obtain the electrohydrostatic pressure
between two planar superconducting bodies, we first solve the electrohydrostatic condition sourced by two finitely separated ionic backgrounds. For each separation length \(L \in [0,20\xi]\), we then integrate the resulting electrohydrostatic energy density,
\begin{equation}\label{eq:electrohydrostatic_energy_density}
    \mathcal{H} = \underbrace{\frac{\epsilon_0}{2}\abs{\frac{\hbar^2}{2mq}\grad\frac{\laplacian\sqrt{n}}{\sqrt{n}}}^2}_{u_\text{electric}} + \underbrace{\frac{\hbar^2}{8m} n\abs{\grad \ln{n}}^2\vphantom{\bigg|}}_{u_\text{elastic}},
\end{equation}
over all space \(V\) and compute the pressure \(P = -\nabla_L \int_V
\mathcal{H}dx\). Details of the calculation are provided in
Fig. \ref{fig:main_fig}, with the main conclusion being the existence
of a negative (attractive) pressure between plates that vanishes in
the limit of zero or infinite separation and reaches a peak for \(L
\approx \xi\). For a conventional (\(m =
2m_e\), \(q = 2e\)) superconductor with \(\lambda_\text{L} \approx 100
\text{ nm}\), the pressure achieves a maximum value
of \(\approx 40 \text{ N/mm}^2\) at separations on the order of
\(1\text{ \AA}\). \PRB{Numerical details are given in Appendix \ref{numerical_details_appendix}, but no explicit calculations are necessary to deduce the pressure's negative sign and the relevant length scale. The former is a consequence of the electrohydrostatic energy density (Eq.~\eqref{eq:electrohydrostatic_energy_density}) being positive semi-definite and therefore minimized by the uniform solution, and the latter is a consequence of the healing length (Eq.~\eqref{eq:healing_length}) emerging as the only length scale in the electrohydrostatic condition (Eq.~\eqref{eq:electrostatics}).}

\PRB{\subsection{Surface energy}}
\PRB{The literature currently lacks experimental data for pressure measurements between superconductors at the angstrom scale. In the absence of such measurements,} the theory's validity can be indirectly assessed by comparing the resulting surface energy,
\begin{equation} \label{eq:surface_energy}
    \sigma_\text{sc} = \frac{1}{2}\int_V\mathcal{H}dx\Big\rvert_{L=0}^{L=\infty} \approx 0.21\xi u_0,
\end{equation}
with existing experimental values \cite{vitos1998surface}. The surface energy represents the energy required, per unit area of new surface formed, to split a material in two along a plane~\cite{lang1970theory}. It is most often measured in the liquid phase, where the observed temperature dependence is extrapolated to yield a value at zero temperature. Prior justifications for this crude approximation exploit simple models of the surface energy's temperature dependence \cite{huang1949surface} and produce a figure of merit commonly used in the assessment of fermionic models for normal metals \cite{lang1970theory, vitos1998surface}. In Table \ref{table:surface_energies}, we provide the surface energies predicted by our model for a number of elemental superconductors, along with experimental values, and find the two to be on the same order of magnitude. For comparison, we also include the surface energy predicted by the HF method applied to the normal fermionic state, which \PRB{yields negative values} and therefore fails at high densities~\cite{lang1970theory}. The tabulated predictions for normal metals are derived from a Jellium background and thus represent the closest analog to this work. Normal metal surface energies have been predicted with greater accuracy via resolution of the ionic background's lattice periodicity, which suggests similar improvements can be made to our model. 

\PRB{\subsection{Screening}}
Eq. \eqref{eq:electrostatics} evidently implies a screening response that differs considerably from that of a normal metal. A direct comparison is most effectively achieved via the Green's function. As derived in \PRB{Appendix \ref{biharmonic_appendix}}, for source distributions representing small perturbations from a uniform background, the electrohydrostatic condition reduces to a self-sourced version of the inhomogeneous biharmonic equation arising in linear elasticity theory~\cite{elasticity},
\begin{subequations}
    \begin{equation}
        n_\text{src} \equiv n_\text{s} + \delta n_\text{src},~\abs{\delta n_\text{src}} \ll n_\text{s}
    \end{equation}
    \begin{equation} \label{eq:linear_elasticity}
        \left(1 + \xi^4\nabla^4\right)\delta n \approx \delta n_\text{src}
    \end{equation}
    \begin{equation} \label{eq:green_SC}
        G\left(\vec{x},\vec{x}'\right) = \frac{1}{4\pi \abs{\vec{x}-\vec{x}'} \xi^2}e^{-\frac{\abs{\vec{x}-\vec{x}'}}{\xi\sqrt{2}}}\sin\left(\frac{\abs{\vec{x}-\vec{x}'}}{\xi\sqrt{2}}\right),
    \end{equation}
\end{subequations}
with \(\delta n \equiv n - n_\text{s}\) the first order perturbation in the superconducting number density, \(\nabla^4 \equiv \laplacian\laplacian\) the biharmonic operator, and \(G\left(\vec{x},\vec{x}'\right)\) the ensuing Green's function. The screening response exhibits both decaying and oscillatory behavior on the length scale of the healing length, which stands in contrast to the Yukawa-like response arising from \PRB{TF} screening in normal metals~\cite{ashcroft2022solid}. The TF theory describes the response of the electron number density \(n_\text{TF}\) to small perturbations from a uniform background. For metals with bulk electron number density \(n_\text{e}\),
\begin{subequations}
    \begin{equation}
        n_\text{src} \equiv n_\text{e} + \delta n_\text{src},~\abs{\delta n_\text{src}} \ll n_\text{e}
    \end{equation}
    \begin{equation}
        \left(1 - \xi_\text{TF}^2 \laplacian\right)\delta n_\text{TF} \approx \delta n_\text{src}
    \end{equation}
    \begin{equation}
        G_\text{TF}\left(\vec{x}, \vec{x}'\right) = \frac{1}{4\pi \abs{\vec{x}-\vec{x}'} \xi_\text{TF}^2} e^{-\frac{\abs{\vec{x}-\vec{x}'}}{\xi_\text{TF}}},
    \end{equation}
\end{subequations}
with \(\delta n_\text{TF} \equiv n_\text{TF} - n_\text{e}\) the first order perturbation in the normal electron number density, \(G_\text{TF}\left(\vec{x},\vec{x}'\right)\) the ensuing Green's function, and \(\xi_\text{TF} \approx 0.34\sqrt{r_\text{s}/a_0}\)~\AA~the TF screening length. In this definition, \(r_\text{s}\) and \(a_0\) are the Wigner-Seitz and Bohr radii, respectively. \PRB{We note that while our healing length and the TF screening length scale differently with the mobile charge carrier density (\(\xi \sim n_\text{s}^{-1/4}\), \(\xi_\text{TF} \sim n_\text{e}^{-1/6}\))}, they \PRB{take on} remarkably similar values for the \PRB{elemental} superconductors listed in Table \ref{table:surface_energies}. \PRB{We emphasize that the two responses are not mutually exclusive:} TF screening by unpaired electrons is a separate and complementary phenomenon that can be incorporated into our model by an appropriate choice of the free current \cite{gabadadze2009effective, gabadadze2008charged}. Here, we have assumed a static Jellium background that neglects TF screening, since the mobile electrons are predominantly paired far below the critical temperature. This is opposite to the GL theory's typical assumption of a TF screening response near the critical temperature, where the mobile electrons are predominantly unpaired. We conclude our discussion of the screening response by noting that the oscillations in Eq. \eqref{eq:green_SC} are showcased in Fig. \ref{fig:main_fig}, where increases (decreases) in electric energy arise from constructive (destructive) interference of screening charges, thereby producing a zero-crossing in the electric pressure. Derivation of the analogous Friedel oscillations in normal metals requires more complex models than TF screening, such as Lindhard theory~\cite{ashcroft2022solid}, which highlights the efficacy of our comparatively simple model.

\PRB{\section{Outlook}\label{outlook}}
Though some confirmation of our model can be found from experimental values for the surface energy, its validity is most directly assessed by experimental measurements of the pressure or screening response. The viability of such measurements requires knowledge of  other forces present at this scale, such as the Casimir and van der Waals interactions~\cite{RevModPhys.88.045003}. \PRB{While the literature contains both theoretical \cite{bimonte2019casimir} and experimental \cite{PhysRevLett.121.030405} analyses that suggest the Casimir force at large separations is unaffected by the superconducting phase transition, we are unaware of any conclusive evidence that suggests this correspondence holds for the small separations considered here. Macroscopic treatments of Casimir forces often rely on assumptions (e.g., local response~\cite{RevModPhys.81.1827}) that break down at and below nanometric gaps. A more detailed analysis incorporating the weaker superconducting response along with nonlocal/atomistic effects is therefore needed and left to future work. Indeed, due to their distinct screening responses, superconductors may exhibit a nonlocal damping of the Casimir force at small separations that differs considerably from its normal metal counterpart.}
Nonetheless, the literature contains some arguments based on UV cutoffs that suggest a Casimir contribution to the surface energy that is comparable to experimental values~\cite{schwinger1978casimir}, \PRB{thereby leaving the significance of the predicted electrohydrostatic contribution uncompromised.} We conclude by noting that since \(C^4\) continuity of the superconducting number density is guaranteed by the nonlocal quantum potential~\cite{bohm1975intuitive}, all contributions to the electrohydrostatic energy density are, in contrast to the Casimir force, finite~\cite{casimir1948attraction}.

By virtue of its formulation in terms of hybridized radiation-matter fields, our model may also be applied to the study of Josephson dynamics, as has been studied in~\cite{dec-qed}. For the static limit considered here, such effects vanish. \PRB{Neglecting the effect of these limit cycle solutions on the pressure between superconducting bodies can be justified by the assumption that the ground state is electrostatic.
We leave the proof of this intuitively plausible assumption to future work but note that its validity would imply that these semiclassical excitations are frozen out in the zero temperature limit, thereby leaving the predicted pressure unaffected in this regime.}

To summarize the results of this study, we have presented a theory of superconductivity akin to the GL theory that is capable of describing the dynamics of superconducting quantum devices well below the critical temperature, and we have used the theory to predict a negative electrohydrostatic pressure between superconducting bodies. Moreover, we have calculated the resulting surface energies and shown that they are in better agreement with experimental values than those predicted by the HF method applied to the normal fermionic state. The screening response predicted by our model is functionally distinct from the TF screening response of normal metals, with a screening length that exhibits a different \PRB{scaling with} the mobile charge carrier density. Nonetheless, the two \PRB{electric-field} screening lengths \PRB{take on} remarkably similar values for the \PRB{elemental} superconductors listed in Table \ref{table:surface_energies}. This agreement suggests that the common assumption of an identical electrostatic response in normal metals and superconductors may originate from the experimental misidentification of our proposed healing length with the TF screening length. A measurement of the electrohydrostatic pressure capable of determining the theory's validity may thus depend on a material choice for which these two screening lengths are sufficiently different. \PRB{Like the GL theory, our model may be applied to the analysis of conventional and unconventional superconductors alike, so material choices need not be restricted to the former. The viability of a pressure measurement also requires an account of the other relevant} forces present at this scale, such as the Casimir and van der Waals interactions~\cite{RevModPhys.88.045003}, which remains an open problem. \PRB{Though not necessary for a pressure measurement, our model may also be applied to the analysis of dynamical phenomena, such as the Josephson effect.}

The authors thank Wentao Fan, Zoe Zager, and Terry Orlando for insightful discussions. This material is based upon work supported by the National Science Foundation Graduate Research Fellowship under Grant No. DGE-2039656, by the US Department of Energy,
Office of Basic Energy Sciences, Division of Materials
Sciences and Engineering, under Award No. DESC0016011, by the National Science
Foundation under the Emerging Frontiers in Research
and Innovation (EFRI) program, Award No. EFMA164098, by the Defense Advanced Research Projects Agency
(DARPA) under Agreements No. HR00111820046, No.
HR00112090011, and No. HR0011047197, and by a
Princeton SEAS Innovation Grant. \\

\appendix
\section{Electrohydrodynamic representation of the Hamiltonian}\label{electrohydrodynamic_hamiltonian_appendix}
The following derivation assumes electrostatic sources \(\vec{j} = \vec{0}\):
\begin{widetext}
\begin{equation} \label{eq:electrohydrodynamic_hamiltonian}
    \begin{aligned}
        \int\mathcal{H}d^3\vec{x} &= \int\left(\frac{\partial \mathcal{L}}{\partial \dot{A}_\rho}\dot{A}_\rho + \frac{\partial \mathcal{L}}{\partial \dot{\psi}}\dot{\psi} + \frac{\partial \mathcal{L}}{\partial \dot{\psi^*}}\dot{\psi^*} - \mathcal{L}\right)d^3\vec{x}\\
        &= \int\left(-\frac{1}{c\mu_0} F^{0\rho}\dot{A}_\rho + i\hbar\psi^*\dot{\psi} - \mathcal{L}\right)d^3\vec{x}\\
        &= \int\left(-\frac{1}{c\mu_0} F^{0\rho}\dot{A}_\rho + \left(j_0 + cq\abs{\psi}^2\right)A_0 + \frac{1}{4\mu_0}F^{\mu\nu}F_{\mu\nu}  + \frac{1}{2m}\psi^*\left(\frac{\hbar}{i}\grad - q\vec{A}\right)^2\psi \right)d^3\vec{x}\\
        &= \int\left(\frac{\epsilon_0}{2}\abs{\vec{E}}^2 + \frac{1}{2\mu_0}\abs{\vec{B}}^2 + \frac{1}{2m}\abs{\left(\frac{\hbar}{i}\grad - q\vec{A}\right)\psi}^2 \right)d^3\vec{x}\\
        &= \int\left(\frac{\epsilon_0}{2}\abs{\vec{E}}^2 + \frac{1}{2\mu_0}\abs{\vec{B}}^2 + \frac{1}{2m}\abs{\sqrt{n}\left(\hbar\grad\theta - q\vec{A}\right) + \frac{\hbar}{i}\grad\sqrt{n}}^2 \right)d^3\vec{x}\\
        &= \int\left(\frac{\epsilon_0}{2}\abs{\vec{E}}^2 + \frac{1}{2\mu_0}\abs{\vec{B}}^2 + n\left(\frac{1}{2}mv^2\right) + \frac{\hbar^2}{8m}n\abs{\grad\ln{n}}^2\right)d^3\vec{x}.\\
    \end{aligned}
\end{equation}
\end{widetext}
    To arrive at the fourth line in Eq. \eqref{eq:electrohydrodynamic_hamiltonian}, we have employed Gauss's law and integration by parts. For more general source distributions, an analog of Poynting's theorem can be derived directly from the equations of motion, 
    \begin{equation} \label{eq:energy_conservation}
        \dot{\mathcal{H}} + \div\vec{S} + \vec{j}\cdot\vec{E} = 0,
    \end{equation}
    where the directional energy flux in electrohydrodynamic form is given by
    \begin{equation}
        \vec{S} \equiv \frac{1}{\mu_0}\vec{E}\cross\vec{B} + n\vec{v}\left(\frac{1}{2}mv^2 - \frac{\hbar^2}{2m}\frac{\laplacian\sqrt{n}}{\sqrt{n}}\right) - \frac{\hbar^2}{4m }\dot{n}\grad \ln{n}.
    \end{equation}
    
\section{Electrohydrostatic condition}\label{electrohydrostatic_condition_appendix}
    The equations of motion are invariant under the gauge transformation \mbox{\(\left({A}^\mu, \theta\right) \rightarrow \left({A}^\mu + \partial^\mu f, \theta - \frac{q}{\hbar}f\right)\)} for any single-valued smooth function \(f\), which motivates us to define the gauge-invariant four-potential \mbox{\(\mathcal{A}^\mu \equiv A^\mu + \frac{\hbar}{q}\partial^\mu\theta\)}. In terms of these variables, Maxwell's forms are preserved. Namely, the electric and magnetic fields are given by \(\vec{E} = -c\grad{\mathcal{A}_0}-\dot{\mathbfcal{A}}\) and \(\vec{B} = \curl\mathbfcal{A}\), respectively, and the electromagnetic tensor is given by \(F^{\mu\nu} = \partial^\mu\mathcal{A}^\nu - \partial^\nu\mathcal{A}^\mu\), which for notational consistency we now refer to as \(\mathcal{F}^{\mu\nu} \equiv F^{\mu\nu}\). In terms of the gauge-invariant four-potential, Maxwell's equations thus undergo a trivial relabeling:
    \begin{equation}
        ~~~~\partial_\mu F ^{\mu\nu} = \mu_0\left(\mathcal{J}^\nu + j^\nu\right) \implies \partial_\mu\mathcal{F}^{\mu\nu} = \mu_0\left(\mathcal{J}^\nu + j^\nu\right).
    \end{equation}
    Since the flow velocity \(\vec{v} = -(q/m)\mathbfcal{A}\), the superconducting four current may be written purely in terms of the number density and the gauge-invariant four-potential
    as \(\mathcal{J}^\mu = qn\left(c, -(q/m)\mathbfcal{A}\right)^\mu\). We proceed by expressing the imaginary and real parts of Schrödinger's equation in polar form in terms of the gauge-invariant four-potential, which correspond to the superconducting charge continuity equation and the quantum Hamilton-Jacobi equation, respectively:
    \begin{widetext}
    \begin{subequations} \label{eq:madelung}
        \begin{align}
            \partial_\mu \mathcal{J}^\mu = 0 &\implies \dot{n} = \frac{q}{m}\div\left(n\mathbfcal{A}\right)\\
            -\hbar\dot\theta = \frac{1}{2}mv^2 - \frac{\hbar^2}{2m}\frac{\laplacian\sqrt{n}}{\sqrt{n}} + qcA_0 &\implies
            -qc\mathcal{A}_0 = \frac{q^2}{2m}\abs{\mathbfcal{A}}^2 - \frac{\hbar^2}{2m}\frac{\laplacian\sqrt{n}}{\sqrt{n}}.
    \end{align}
    \end{subequations}
    \end{widetext}
    To ground the reader, we note that upon taking the gradient of the quantum Hamilton-Jacobi equation, Eqs. \eqref{eq:madelung} are simply the Madelung equations. In an electrostatic system (\(\vec{j} = \mathbfcal{A} = \vec{0}\)), the only non-trivial component of Maxwell's equations is Gauss's law, and the only non-trivial component of Schrödinger's equation is the quantum Hamiltonian-Jacobi equation, which are respectively given by
    \begin{subequations}
        \begin{align}
            \laplacian \left(c\mathcal{A}_0\right) & = -\frac{1}{\epsilon_0}\left(qn+cj_0\right)\\
            c\mathcal{A}_0 &= \frac{\hbar^2}{2mq}\frac{\laplacian\sqrt{n}}{\sqrt{n}}.
        \end{align}
    \end{subequations}
    Combining the two equations to eliminate \(\mathcal{A}_0\), we arrive at the electrohydrostatic condition,
    \begin{equation}
        \frac{\hbar^2}{2mq}\laplacian{\frac{\laplacian\sqrt{n}}{\sqrt{n}}} = -\frac{1}{\epsilon_0}\left(qn + cj_0\right),
    \end{equation}
    which can be expressed in terms of the normalized number densities as given in the main text. We note that the electric field may be written purely in terms of the number density by taking the negative gradient of the quantum Hamilton-Jacobi equation \mbox{\(\vec{E} = -\hbar^2/(2mq)\grad\left(\laplacian\sqrt{n}/\sqrt{n}\right)\)}.

\section{Numerical details}\label{numerical_details_appendix}
    The non-dimensionalized 1D electrohydrostatic condition and energy density employed in this study are given by
    \begin{subequations}
        \begin{gather}
            \bar{n} + 2\frac{d^2}{d\tilde{x}^2}\left(\frac{1}{\sqrt{\bar{n}}}\frac{d^2\sqrt{\bar{n}}}{d\tilde{x}^2}\right) = \Theta\left(\abs{\tilde{x} - \tilde{L}/2}\right)
            \\
            \int{\mathcal{H}dx} = u_0\xi\int\Bigg(\underbrace{\abs{\frac{d}{d\tilde{x}}\left(\frac{1}{\sqrt{\bar{n}}}\frac{d^2\sqrt{\bar{n}}}{d\tilde{x}^2}\right)}^2}_{u_\text{electric}~[u_0]} + \underbrace{\frac{\bar{n}}{4} \abs{\frac{d\ln{\bar{n}}}{d\tilde{x}}}^2}_{u_\text{elastic}~[u_0]}\Bigg)d\tilde{x},
        \end{gather}
    \end{subequations}
    where \(\tilde{x} \equiv x/\xi\), \(\tilde{L} \equiv L/\xi\), and \(\Theta\) is the Heaviside theta function. Calculation of the electrohydrostatic pressure between superconducting bodies was performed numerically using a central finite difference scheme with second-order accuracy for all spatial derivatives. The electrohydrostatic condition was solved on a numerical grid consisting of 801 equally spaced points \(\tilde{x} \in [-20,20]\) with boundary conditions \(\bar{n}(\tilde{x})\rvert_{\abs{\tilde{x}}\geq20} = 1\) for 201 equally spaced separation lengths \(\tilde{L} \in [0,20]\). Analytic manipulations of the electrohydrostatic condition were performed with SymPy, and the ensuing nonlinear vector equation was solved numerically using \verb|scipy.optimize.fsolve|, which converged with the default tolerance of \verb|1.49012e-08| for all results in this study.

\section{Self-sourced inhomogeneous biharmonic equation and Green's function}\label{biharmonic_appendix}
    For a uniform background \(\bar{n}_\text{src} = 1\), the electrohydrostatic condition (Eq.~\eqref{eq:electrostatics}) yields the trivial solution \(\bar{n} = 1\). We derive the self-sourced version of the inhomogeneous biharmonic equation arising in linear elasticity theory by considering source distributions representing small perturbations from a uniform background as follows,
    \begin{subequations}
    \begin{align}
        \bar{n}_\text{src}\ofx &= 1 + \lambda\bar{n}_\text{src}^{(1)}\ofx \\
        \bar{n}\ofx &= 1 + \sum_{k=1}^\infty \lambda^k \bar{n}^{(k)}\ofx,
    \end{align}
    \end{subequations}
    so that as \(\lambda \rightarrow 0\), we recover the unperturbed uniform medium. To first order in \(\lambda\), Eq. \eqref{eq:electrostatics} reads
    \begin{equation} \label{eq:electrostatic_first_order}
        \begin{aligned}
            \lambda\bar{n}^{(1)} + \frac{\partial}{\partial \lambda}\left(2\xi^4\laplacian\frac{\laplacian \sqrt{\bar{n}}}{\sqrt{\bar{n}}}\right)\Bigg\rvert_{\lambda = 0} \lambda &= \lambda\bar{n}_\text{src}^{(1)}\\
            \bar{n}^{(1)} + \xi^4 \nabla^4 \bar{n}^{(1)} &= \bar{n}^{(1)}_\text{src},
        \end{aligned}
    \end{equation}
    as given in the main text with \(\delta\bar{n}_\text{src} \equiv \lambda\bar{n}^{(1)}_\text{src}\) and \(\delta\bar{n} \equiv \lambda\bar{n}^{(1)}\).
    To arrive at the second line, we have employed the quantum potential derivative identity derived in Eq.~\eqref{eq:pressure_identity}. We now derive the corresponding Green's function by considering a source distribution of the form \(\bar{n}^{(1)}_\text{src}(\vec{x}) = \delta^{(3)}\ofx\).
    Expanding \(\bar{n}^{(1)}\) and \(\bar{n}^{(1)}_\text{src}\) in the Fourier basis,
    \begin{subequations}
        \begin{align} 
            \bar{n}^{(1)}\ofx&\equiv \int \tilde{n}^{(1)}(\vec{k})e^{i\vec{k}\cdot\vec{x}}d^3\vec{k}\\
            \bar{n}_\text{src}^{(1)}\ofx &\equiv \int \tilde{n}_\text{src}^{(1)}(\vec{k})e^{i\vec{k}\cdot\vec{x}} d^3\vec{k},
        \end{align}
    \end{subequations}
    and taking the inverse Fourier transform of Eq. \eqref{eq:electrostatic_first_order} yields
    \begin{equation}
        \left(1 + (k\xi)^4\right)\tilde{n}^{(1)}(\vec{k}) = \tilde{n}_\text{src}^{(1)}(\vec{k}) = \frac{1}{(2\pi)^3}
    \end{equation}
    with \(k \equiv \abs{\vec{k}}\). We proceed by solving for \(\bar{n}^{(1)}\) in spherical coordinates with \(\rho \equiv \abs{\vec{x}}\):
    \begin{equation}
        \begin{aligned}
            \bar{n}^{(1)}\ofx &= \frac{1}{(2\pi)^3}\int \frac{e^{i\vec{k}\cdot\vec{x}}}{1+(k\xi)^4}d^3\vec{k}\\
            &= \frac{1}{(2\pi)^3}\int_0^\infty \int_0^\pi\int_0^{2\pi} \left(\frac{e^{ik\rho\cos{\theta}}}{1+(k\xi)^4}k^2\sin\theta \right)d\phi d\theta dk\\
            &= \frac{1}{(2\pi)^2}\int_0^\infty \int_0^\pi\left(\frac{e^{ik\rho\cos{\theta}}}{1+(k\xi)^4}k^2\sin\theta \right) d\theta dk\\
            &= \frac{1}{2\pi^2\rho}\int_0^\infty \left(\frac{k\sin(k\rho)}{1+(k\xi)^4}\right) dk\\
            &= \frac{1}{4\pi \xi^2\rho} \sin\left(\frac{\rho}{\xi\sqrt{2}}\right)e^{-\frac{\rho}{\xi\sqrt{2}}}.
        \end{aligned}
    \end{equation}
    
    For source distributions of the form \(\bar{n}^{(1)}_\text{src}(\vec{x}) = \delta^{(3)}(\vec{x}-\vec{x}')\), the first order response \(\bar{n}^{(1)}(\vec{x})\) can be attained by a simple coordinate shift, which yields the Green's function in the main text. We note that the total net charge is neutral, as ensured by the property \(\int_V\bar{n}^{(1)}d^3\vec{x} = \int_V\bar{n}_\text{src}^{(1)}d^3\vec{x}\), where integration is performed over all space \(V\). \\

\section{Quantum potential derivative identity}
    The derivative of the quantum potential with respect to an arbitrary variable \(\alpha\) can be written as follows,
            \begin{equation} \label{eq:pressure_identity}
                \begin{aligned}
                    \frac{\partial}{\partial \alpha}\frac{\laplacian\sqrt{n}}{\sqrt{n}}
                    &= \frac{1}{2}\left(\frac{\laplacian\frac{n'}{\sqrt{n}}}{\sqrt{n}} - \frac{\laplacian\sqrt{n}}{n^{3/2}}n'\right)\\
                    &= \frac{1}{2n^3}\left(n^2\laplacian - n\grad n \cdot \grad + \abs{\grad n}^2 - n\laplacian n\right)n'\\
                    &= \frac{1}{2n^3}\commutator{n}{\grad}^2 \frac{\partial n}{\partial \alpha},
                \end{aligned}
            \end{equation}
        where \(n' \equiv \partial n / \partial\alpha\), \(\commutator{n}{\grad}f \equiv (n\grad)f - (\grad n)f\), and we have employed the dyadic notation \(\grad\vec{f} \equiv \div\vec{f}\).
        
\bibliography{refs}

\end{document}